\documentclass[epj]{webofc}
\usepackage[varg]{txfonts}   
\usepackage{graphicx,color} 
\usepackage{bm}       
\usepackage{amsmath}  
\usepackage{amssymb}
\woctitle{21st International Conference on Few-Body Problems in Physics}
\begin{document}
\title{Experiments towards resolving the proton charge radius puzzle}
\author{A. Antognini\inst{1,2}\fnsep\thanks{\email{aldo@phys.ethz.ch}} \and
K. Schuhmann\inst{1,2}       \and
F. D. Amaro\inst{3}         \and
P. Amaro\inst{4}               \and
M. Abdou-Ahmed\inst{5}        \and
F. Biraben\inst{6}          \and
T.-L. Chen\inst{7}            \and
D. S. Covita\inst{8}          \and
A. J. Dax\inst{2}               \and
M. Diepold\inst{9}                 \and
L. M. P. Fernandes\inst{3}      \and
B. Franke\inst{9}            \and
S. Galtier\inst{6}             \and
A. L. Gouvea\inst{3}              \and
J. G\"otzfried\inst{9}         \and
T. Graf\inst{5}                 \and
T. W. H\"ansch\inst{9}          \and\thanks{Also at: Ludwig-Maximilians-Universit\"at, 80539 Munich, Germany.}
M. Hildebrandt\inst{2}          \and
P. Indelicato\inst{6}              \and
L. Julien\inst{6}                \and
K. Kirch\inst{1,2}              \and
A. Knecht\inst{2}             \and
F. Kottmann\inst{1}             \and
J. J. Krauth\inst{9}         \and
Y.-W. Liu\inst{7}                  \and
J. Machado\inst{4,6}            \and
C. M. B. Monteiro\inst{3}       \and
F. Mulhauser\inst{9}         \and
F. Nez\inst{6}                \and
J. P. Santos\inst{4}             \and
J. M. F. dos Santos\inst{3}      \and
C. I. Szabo\inst{6}              \and
D. Taqqu\inst{1}                 \and
J. F. C. A. Veloso\inst{8}       \and
A. Voss\inst{5}                  \and
B. Weichelt\inst{5}              \and
R. Pohl\inst{9}                 
}

\institute{
Institute for Particle Physics, ETH, 8093 Zurich, Switzerland
\and
Paul Scherrer Institute, 5232 Villigen-PSI, Switzerland
\and
 LIBPhys, Physics Department, Universidade de Coimbra, 3004-516 Coimbra, Portugal
\and
Laborat\'{o}rio de Instrumenta\c{c}\~{a}o, Engenharia Biom\'{e}dica e F{\'\i}sica da
Radia\c{c}\~{a}o (LIBPhys-UNL) e Departamento de F{\'\i}sica da Faculdade de
Ci\^encias e Tecnologia da Universidade Nova de Lisboa, Monte da Caparica, 2892-516
Caparica, Portugal.
\and
Institut f\"ur Strahlwerkzeuge, Universit\"at Stuttgart, 70569 Stuttgart, Germany
\and
Laboratoire Kastler Brossel, UPMC and CNRS, 75005 Paris, France
\and
Physics Department, National Tsing Hua University, Hsincho 300, Taiwan
\and
i3N, Universidade de Aveiro, Campus de Santiago, 3810-193 Aveiro, Portugal
\and
Max Planck Institute of Quantum Optics, 85748 Garching, Germany
          }

\abstract{We review the status of the proton charge radius
  puzzle. Emphasis is given to the various experiments
  initiated to resolve the conflict between
  the muonic hydrogen results and the results from scattering and
  regular hydrogen spectroscopy.
}
\maketitle
%

\section{The proton charge  radius puzzle}
\label{sec-1a}

The historical route to the proton charge radius $(r_p)$ is from
elastic electron-proton scattering.
In a completely complementary fashion, it has been obtained also from
``high-precision'' laser spectroscopy of hydrogen (H).
Since a few years,  ``high-sensitivity'' laser spectroscopy of muonic
hydrogen ($\mu$p) offers a third way.
The value extracted from $\mu$p with a relative accuracy of $5\times
10^{-4}$ is an order of magnitude more accurate than obtained from the
other methods.
Yet the value is 4\% smaller than derived from electron-proton
scattering and H spectroscopy with a disagreement at the $7\sigma$
level~\cite{Pohl2010, Antognini2013, deBeauvoir2000, sick:2014:fewbody,
  Bernauer2012}.

In the last five years as summarized in~\cite{Pohl2013,Carlson2015}
various cross checks and refinements of bound-state QED calculations
needed for the extraction of $r_p$ from $\mu$p have been performed,
together with investigations of the proton structure.
Several suggestions in the field of ``beyond standard model'' BSM physics have
been articulated, re-analysis of scattering data have been carried out
and new experiments have been initiated.
Despite this, presently the discrepancy still persists and the resolution
of the proton radius puzzle remains unknown.
In this article, we summarize mainly the ongoing experimental
activities which hold the potential to unravel the proton radius
puzzle.

As the atomic energy level are slightly modified by the nuclear finite
size, it is possible to deduce the nuclear charge radius by performing
spectroscopy of the atomic energy levels.
In leading order, the energy shift caused by the nuclear finite size is 
\begin{equation}
\Delta E_\mathrm{finite \;size}= \frac{2\pi \alpha}{3} |\phi^2(0)|^2 R_E^2
                      =  \frac{2 m_r^3 \alpha^4}{3n^3}R_E^2
\label{eq:finite-size}
\end{equation}
where $\phi(0)$ is the wavefunction at the origin in coordinate space,
$m_r$ the reduced mass, $\alpha$ the fine structure constant and $n$
the principal quantum number.
$R_E$ is the charge radius of the nucleus  defined in a
covariant way as the slope of the electric form factor ($G_E$) at zero
momentum exchange $Q^2$
\begin{equation}
R_E=-6\frac{dG_E}{dQ^2}\Big|_{Q^2=0} \,.
\label{eq:def}
\end{equation}
Non-relativistically,  $R_E$ is  the
second moment of the electric charge distribution $\rho_E$ of the
nucleus
%
$R_E^2\approx\int \vec{dr}\, \rho_E({\vec r})r^2$.
%

The $m_r^3$ dependence of Eq.~(\ref{eq:finite-size}) reveals  the
advantages related with muonic atoms.
As the muon mass is 200 times larger than the electron mass, the
muonic wavefunction strongly overlaps with the nucleus ensuing a large
shift of the energy levels due to the nuclear finite size.
Thus, the muonic bound-states represent ideal systems for the precise
determination of nuclear charge radii.

The proton form factors can be obtained from unpolarized differential
cross section measurements of electron scattering off protons.
In the one-photon approximation the elastic cross
section is
\begin{equation}
\frac{d\sigma}{d\Omega}=\frac{d\sigma}{d\Omega}\Big|_{Mott} \times \frac{1}{1+\tau}\Big( G^2_E(Q^2)+\frac{\tau}{\epsilon}G^2_M(Q^2)  \Big)
\end{equation}
where the Mott cross section applies for point-like
particles, $G_E$ and $G_M$ are  the electric and magnetic form factors of
the proton, $\tau=Q^2/4M^2$ and $\epsilon^{-1}=1+2(1+\tau)
\tan^2{(\theta/2)}$ are kinematical variables.
%
Commonly, the Rosenbluth separation is applied to disentangle the
charge from the magnetic contributions by using the angle-dependence
at fixed $Q^2$.
Therefore, by measuring the differential cross section at various
$Q^2$ and angles $\theta$, one obtains $G_E(Q^2)$
and $G_M(Q^2)$ and via Eq.~(\ref{eq:def}) the radius.

\section{Muonic hydrogen and possible new physics explanations}
\label{sec-1a}

The CREMA collaboration has measured two transition in $\mu$p: from
the triplet ($2S_{1/2}^{F=1}-2P_{3/2}^{F=2}$)~\cite{Pohl2010} and the
singlet ($2S_{1/2}^{F=0}-2P_{3/2}^{F=1}$)~\cite{Antognini2013}
2S-states yielding a radius of $r_p=0.84087(39)$~fm.
More specifically the two measured energy splittings, from the triplet
$h\nu_t$ and from the singlet $h\nu_s$ states, can be combined to
obtain both the 2S Lamb shift $E_L=\Delta E(2S-2P_{1/2})$ and the
hyperfine splitting $E_\mathrm{HFS}$:
\begin{eqnarray}
E_L & = &\frac{1}{4} h \nu_s+\frac{3}{4} h \nu_t - 8.8123(2) ~\mathrm{meV},\\
E_\mathrm{HFS} & = & h \nu_s- h \nu_t + 3.2480(2) ~\mathrm{meV},
\end{eqnarray}
where the numerical values follow from reliable  proton-independent
corrections of the 2P states.
Experimentally~\cite{Antognini2013}  
\begin{equation}
E_L^\mathrm{exp}=202.3706(23)~\mathrm{ meV}=  48932.99(55)~\mathrm{GHz}
\end{equation}
limited by statistics while the systematics effects is at the
300~MHz level.  From theory~\cite{Antognini2013_annals}
\begin{equation}
E_L^\mathrm{th}=206.0336(15)-5.2275(10) r_p^2+0.0332(20)~\mathrm{ meV}
\label{eq:lamb-shift-mup}
\end{equation}
where $r_p$ is given in fm. 
The first term accounts for QED contributions, the second for finite
size effects, and the third for the two-photon exchange (TPE)
contribution.
From the Lamb shift an improved $r_p$ value free from uncertainties
related with the HFS splitting has been determined, and from the HFS
the Zemach radius, albeit not with the same accuracy as the charge
radius.

The consistency of the two $\mu$p measurements, represents an
important cross check of the muonic results.
The typical systematic effects affecting the atomic energy levels are
substantially suppressed in $\mu$p due to the stronger binding.
The internal fields and the level separation of the muonic atoms are
greatly enhanced compared to regular atoms making them insensitive to
external fields (AC and DC Stark, Zeeman, black-body radiation and
pressure shifts).
Thus $\mu$p turns out to be very sensitive to the proton charge radius
($m_r^3$-dependence) and insensitive to systematics which typically  scales
as $\sim1/m_r$.
The possible involvement of weakly bound three-body systems in the
muonic hydrogen spectroscopy experiment~\cite{Jentschura2011} has been
ruled out by three-body calculations~\cite{Karr2012}, and by the
experimental non-observation of sizable additional line broadening,
line splitting and event rate decrease.

The bound-state QED corrections which give rise to
Eq.~(\ref{eq:lamb-shift-mup}) have been computed by several
groups~\cite{Pachucki1999, Boriev7} as summarized
in~\cite{Antognini2013_annals} and updated recently
in~\cite{Karshenboim2015_th, Peset2015b}.
Particular attention has been devoted to the TPE contribution which
has been computed in two frameworks: one making use of dispersion
relations and measured inelastic structure functions of the
proton~\cite{Carlson2011}, the second from chiral perturbation
theories~\cite{Pineda2014}.
Both ways provide consistent predictions.
In the dispersion-based approach a subtraction term in form of an
integral from zero to infinite $Q^2$ is necessary which
can not be fully constrained by data.
%
At intermediate $Q^2$ indeed a modeling of the proton is
necessary~\cite{Hill2011}.
The large majority of the community, agrees with a value for this
subtraction term which is two order of magnitude smaller than the
measured discrepancy of 0.3~meV~\cite{Birse, Gorchtein, Alarcon}.
Even if improbable, in principle still a very un-smooth and physically
unmotivated proton structure could be constructed which could shift
the $\mu$p transition to explain the measured
discrepancy~\cite{Miller2013}.
However, the published subtraction functions proposed to solve the
proton radius puzzle would affect through the Cottingham formula the
proton and neutron masses by 600~MeV~\cite{Walker2012} which is quite
implausible when compared with the measured and computed
neutron-proton mass difference of 1.29~MeV~\cite{Borsanyi2015}.

Explaining the discrepancy by this subtraction term can be interpreted as
an exotic hadronic effect.
Other effects occurring in the vicinity of the nucleus, as the
breakdown of the perturbative approach in the electron-proton
interaction~\cite{Jentschura2015, Pac14}, the interaction with sea $\mu^+\mu^-$ and
$e^+e^-$ pairs~\cite{Jentschura2013} etc., have been suggested and need to be
further investigated but at moment are not yet conclusive.
Several BSM extensions have been proposed but the vast majority of them have
difficulties to resolve the measured discrepancy without conflicting
with other low energy constraints.
Still some BSM theories can be formulated 
but they require fine-tuning (e.g. cancellation between axial and
vector components), targeted coupling (e.g. preferentially to the
muon or to muon proton) and are problematic to be merged in a gauge
invariant way into the standard model~\cite{Karshenboim2014,
  Carlson2015b, Brax2014}.

\section{Hydrogen experiments}
\label{sec-1c}

In a simplified way, the hydrogen S-state energy levels can be
described by
\begin{equation}
  E(nS)=\frac{R_\infty}{n^2}+\frac{L_{1S}}{n^3}
\label{eq:H}
\end{equation}
where $R_\infty=3.289 \,841 \,960 \,355(19)
\times 10^{15} $~Hz is the Rydberg constant and approximately
\begin{equation}
L_{1S}\simeq 8171.636(4) + 1.5645 R_E^2\;~\mathrm{ MHz}
\label{eq:lamb-shift-H}
\end{equation}
the Lamb shift of the ground state ($r_p$ expressed in fm) given by
bound-state QED calculations.
The different $n$-dependence of the two terms in Eq.~(\ref{eq:H}) permits
the determination of both $R_\infty$ and $r_p$, from at least two
transition frequencies in H, assuming  Eq.~(\ref{eq:lamb-shift-H}).

Being the most precisely measured transition ($u_r=4\times
10^{-15}$)~\cite{Parthey2011, Matveev2013} and because it is showing
the largest sensitivity to the Lamb shift contributions usually the
1S-2S transition is used.
By combining it with a second transition 
measurement, it is possible to determine the proton charge radius.
%
When taken individually, the various $r_p$ values extracted from H
spectroscopy by combining two frequency measurements (2S-4S, 2S-12D,
2S-6S, 2S-6D, 2S-8S, 1S-3S as ``second'' transition) are statistically
compatible with the value from $\mu$p.
Only the value extracted by pairing the 1S-2S and the
2S-8D transitions is showing a $3\sigma$ 
deviation while all the others differ only by $\lesssim 1.5\sigma$.

So the $4.4\sigma$ discrepancy between the proton charge radius from
$\mu$p and H spectroscopy emerges only after an averaging process
(mean square adjustments of all measured transitions) of the various
``individual'' determinations and consequently is less startling than
it looks at first glance.
A small systematic effect common to the H measurements could be
sufficient to explain the deviation between $\mu$p and H results.
This fact becomes even more evident if we consider the frequency
shifts (absolute and normalized to the linewidth) necessary to match
the $r_p$ values from $\mu$p and H as summarized for selected
transitions in Table~\ref{tab:1}.
\begin{table}[tbh]
\caption{\label{tab:1} Transition frequency shift needed to match the $r_p$ values from H and $\mu$p, expressed also
  relative to the stated experimental accuracy $\sigma$, and to the
  transition effective linewidths $\Gamma_\mathrm{eff} $. } \centering
                  \begin{tabular}{l|rrr}
                    \hline 
                    Transition       & Shift rel. to uncertainty & Absolute shift & Shift rel. to effective linewidth\\
                    \hline 
                    $\mu$p(2S-2P)    &  $100 \, \sigma $    & 75 GHz    & $4\, \Gamma_\mathrm{eff}$\\     
                    H(1S-2S)         &  $4'000 \, \sigma $  & 40 kHz    & 40$\, \Gamma_\mathrm{eff}$\\     
                    H(2S-4P)         &  $1.5\, \sigma    $  &  9 kHz    & $7\times 10^{-4} \,\Gamma_\mathrm{eff}$\\     
                    H(2S-2P)         &  $1.5\, \sigma $     &  5 kHz    &  $ 7\times 10^{-4} \,\Gamma_\mathrm{eff}$\\     
                    H(2S-8D)         &  $ 3 \, \sigma $     &  20 kHz   & $ 2\times 10^{-2} \,\Gamma_\mathrm{eff}$\\     
                    H(2S-12D)        &  $ 1 \, \sigma $     &  8 kHz    &  $ 5\times 10^{-3} \,\Gamma_\mathrm{eff}$\\     
                    H(1S-3S)         &  $ 1 \, \sigma $     &  13 kHz   & $ 5\times 10^{-3} \,\Gamma_\mathrm{eff}$\\     
                    \hline 
                   \end{tabular}
\end{table}
Obviously the discrepancy can not be solved by shifting the 1S-2S and the
$\mu$p measurements because it would require displacements
corresponding to $4000\sigma$ and $100\sigma$,
respectively.
Expressing the required frequency shift relative to the
linewidth as in the last column allows to
better recognize some aspects of the experimental challenges. 
For example a  shift of only $7\times 10^{-4} \,\Gamma$ of the
$2S-4P$ transition would be sufficient to explain the
discrepancy.
A control of the systematics  which could distort and shift the
line shape on this level of accuracy is far from being a trivial task.
Well investigated are the large line broadenings owing to
inhomogeneous light shifts which results in profiles with effective
experimental widths much larger than the natural
linewidths~\cite{deBeauvoir2000}.

Another exemplary correction relevant in this context, named quantum
interference, has been brought recently back to
attention~\cite{Horbatsh2010}, and has lead to various reevaluations
of precision experiments~\cite{Amaro2015}.
An atomic transition can be shifted by the presence of a neighboring
line, and this energy shift $\delta E $, as a rule of thumb, amounts
to~\cite{Horbatsh2010} 
$\frac{\delta E }{ \Gamma} \approx \frac{\Gamma }{D} $
where $D$ is the energy difference between the two resonances
and $\Gamma$ the transition linewidth.
Thus, if a transition frequency is aimed with an absolute accuracy of
$\Gamma/x$, then the influence of the neighboring lines with $D \le
x\Gamma$ has to be considered.
The precise evaluation of these quantum interference effects are challenging
because they require solving numerous differential equations describing
the amplitude of the total excitation and detection processes from 
initial to  final state distributions and because it
depends on experimental parameters such as the angular position
and polarization sensitivity of the detectors, the laser intensity,
direction and polarization, the initial population distribution among
states, etc.

Generally speaking, transition frequencies involving states with large
$n$ are more sensitive to systematical effects caused
by external fields. 
Emblematic is for example the $n^7$-dependence of the DC Stark effect.
Motivated by the possibility that minor effects in H could be
responsible for the observed discrepancy, various activities have been
initiated in this field:
\begin{itemize}
\item {\it $2S-4P_{1/2}$ and $2S-4P_{3/2}$ at MPQ
  Garching~\cite{Beyer2013a}:} They are aiming at improving previous
  measurements by a factor of 5 down to an accuracy of few kHz which
  would yield a $r_p$ with less than 2\% accuracy when paired with the
  1S-2S transition.  Preparation of the 2S state by means of optical
  excitation and an almost $4\pi$ Lyman-alpha detection system are key
  elements to control the line pulling due to quantum interference on
  the $1\times 10^{-4}\Gamma$ level of accuracy required.

\item{\it $1S-3S$ transition at LKB and MPQ~\cite{Galtier2014,
    Peters2013}:} The 2010 results from the LKB group delivered the
  second most precise transition frequency measurement in H
  with a total uncertainty of 13~kHz corresponding to a relative
  accuracy of $4.5\times10^{-12}$. The error budget was dominated by
  statistics (12~kHz) and uncertainties in the velocity distribution
  of the atomic beam (3~kHz).  A 1\% accuracy of the proton radius
  will require a measurement of the 1S-3S transition with accuracy of
  about 2~kHz.
To reach this goal, the Paris group is presently pursuing the
measurement of the 1S-3S transition at 205~nm wavelength using cw
spectroscopy along the same line of investigations as in previous
experiment. Special emphasis is devoted to the velocity dependent
systematic effects.
Oppositely, the MPQ group, to circumvent the difficulties related
with the generation of the 205~nm light, has devised an experiment which
uses pico-second frequency comb pulses.

\item{\it $2S-2P$ classical Lamb shift in Toronto~\cite{Vutha2012}:}
The measurement of the 2S-2P energy splitting alone can lead to $r_p$.
Indeed, as this transition frequency does not depend on $R_\infty$ 
there is no need to combine it with a second transition
frequency measurement.
Microwave spectroscopy based on the Ramsey method of separated
oscillatory field is used for this purpose.
A factor of 5 improvement is anticipated, which implies a determination
of $r_p$  to the 0.6\% level.
To reach this ambitious goal the position of the line has to
be determined with 1 part in $10^4$.
\end{itemize}

The ``second'' (beside the 1S-2S transition) transition frequency
measurement in H can be interpreted as a measurement of the Rydberg
constant.
An alternative way to an independent determination of the Rydberg
constant, is to perform optical spectroscopy of H-like ions between
circular Rydberg states where the nuclear size corrections are
basically absent, the QED contributions small, and the linewidths
narrow~\cite{Tan2011}, or via spectroscopy of positronium and
muonium~\cite{Cooke2015}.

\section{Scattering experiments}
\label{sec-1d}

The Mainz A1 collaboration at MAMI  has measured in 2010 1422 precise
relative e-p cross sections in the low-$Q^2$ regime (0.0038~GeV$^2$ to 0.98~GeV$^2$) and a
wide range of beam energy and scattering angles~\cite{Bernauer2012}.
Two spectrometers were moved with overlapping angle settings while a
third spectrometer was kept fixed and used as a luminosity monitor.

As data are available only down to a minimal $Q^2$, extrapolation to
$Q^2=0$ is required to determine the proton charge radius.
Numerous works have been concerned with the issues related with
this extrapolation procedure.
The Mainz group found a satisfactory goodness of fit ($\chi^2=1.14$
for 1422 points) through the use of flexible fitting
functions (splines and polynomials) and the resulting radius
reads~\cite{Bernauer2012}
$r_p=0.879(5)_\mathrm{stat}(4)_\mathrm{syst}(2)_\mathrm{model}(4)_\mathrm{group}$~fm
in agreement with the CODATA06 value of 0.8768(69)~fm based mostly on
atomic measurements.
It is important that the fit function be flexible enough to adequately
reproduce the data, without being so flexible that over-fitting
occurs. 
One solution to this problem is given in~\cite{sick:2014:fewbody, Sick2014}
where the low-$Q^2$ behavior of the form factor is
constrained by using large-$r$ assumption of the charge distribution.
Reanalysis of the world data using these constraints yields
$r_p=0.879(11)$~fm~\cite{Arrington2015a}.
%
%
Another approach to address issues of over- or
under-fitting data is the use of bounded polynomial $z$ expansion
(after conformal mapping) and constraining the expansion coefficients
to decrease ``perturbatively'' with increasing order~\cite{Lee2015}.
%

Conformal fits to the form factors were performed in~\cite{Hill2010}  
yielding $r_p=0.870(23)(12)$~fm  in agreement with the CODATA value.
Another group fitting the 1422 Mainz data points found
$r_p=0.840(15)$~fm~\cite{Lorenz2015} in agreement with the PSI result
but with a $\chi^2=1.4$ (using more flexible functions the same
group found $\chi^2=1.1$).
Noteworthy, the value of $r_p\approx 0.84$~fm from analysis of
scattering data using dispersion relations and vector-mesons dominance
models was obtained by the same group prior to the publication of the
muonic result.
The use of a form factor model (for all $Q^2$-range) based on
dispersion relation and vector-meson dominance introduce rigidity in
the model which results in the larger $\chi^2$.
So tension exists between the use of a physically motivated model
giving a poorer fit or very flexible fit functions without physical
constraints yielding better $\chi^2$.
This tension could arise by inappropriateness of the theoretical model,
insufficient treatment of experimental details, or due to an
underestimation of the scattering cross section uncertainties.

A wide-ranging study of possible systematic effects of the 2010 Mainz
data has been recently reported in~\cite{Lee2015}.
Special attention was devoted to the extrapolation procedure, to
normalization factors needed to smoothly combine the various
spectrometer settings and to refinement of the radiative corrections.
This reanalysis yields $r_p=0.895(20)$~fm. When  applied to the
world data (excluding Mainz 2010) $r_p=0.918(24)$~fm is
found~\cite{Lee2015}.
Even though some inconsistencies between data sets were found, which have
led to the increased uncertainty of the $r_p$  extracted from
scattering, it remains difficult to reconcile the scattering results 
with the muonic results.
The only sure conclusion is that analysis of low-$Q^2$ scattering data
is not simple and remains a matter of discussion.
Because data at still lower $Q^2$ would be beneficial,
two electron-proton experiments  have been initiated: 
\begin{itemize}
\item {\it PRad experiment at Hall B in JLAB~\cite{Gasparian2014}:} This experiment planned
  to operate at $Q^2$ down to $2\times 10^{-4}$~GeV$^2$ aims to obtain
  $r_p$ with sub-percent accuracy. The experiment is based on a
  windowless target and a downstream calorimeter which allows to
  extend the cross sections measurements to smaller scattering angles.
  The need to measure relative cross sections at about 0.2\% level
  requires knowledge of the angle to 10~ $\mu$rad accuracy which
  makes this experiment very challenging.

\item{\it Initial state radiation at MAMI, A1 collaboration~\cite{Mihovilovic2014}:} Making
  use of the initial state radiation techniques, where the initial
  electron momentum is degraded by photon emission, the
  momentum transfer to the proton is reduced.  The scattered electron
  is measured with the usual spectrometers but no information on the
  photon is observed. However, by comparing measurements with Monte
  Carlo simulations accurate form factors can be determined down to
  $Q^2$ of $2\times 10^{-4}$ GeV$^2$. 

\end{itemize}
Other scattering experiments can provide very important informations:
\begin{itemize}
\item{\it MUSE: muon scattering experiment at PSI~\cite{Kohl2014}:} The MUSE
  experiment plans to measure $\mu^+-p$ and $\mu^--p$ as well as
  $e^+-p$ and $e^--p$ scattering down to 0.002~GeV$^2$. By comparing
  negative with positive charges they will have an handle on the
  insidious TPE contribution. Comparison between electron and muon
  cross sections allows the elimination of common systematical
  effects, including some extrapolation uncertainties. Thus, this
  experiment has not only the potential to measure the proton charge
  radius absolute value to 2\% accuracy, but also a possible
  difference between radii extracted from electron and muon scattering
  down to a relative accuracy of 1\%. In this way possible
  muon-specific interactions can be disclosed.

\item{\it Deuteron scattering at Mainz~\cite{DistlerPC}:} New e-d scattering data have
  been collected in Mainz, with the aims to extract a new value of the
  deuteron charge radius and break-up informations which may be used
  to compute the deuteron polarizability contribution in muonic
  deuterium.
\end{itemize}
Of relevance in this context is also the planned measurement at
JLAB~\cite{Myers2014} aiming at the electric form factor of the mirror
nuclei $^3$He and $^3$H to extract their charge radii difference, and
the TREK program at J-PARC that scrutinizes $K$-decays to search for BSM
physics motivated by the $r_p$ puzzle.

\section{Muonic deuterium and muonic helium ions}
\label{sec-2c}

In 2009 the CREMA collaboration measured two 2S-2P transitions in
muonic deuterium ($\mu$d).
The 2S-2P energy splitting in $\mu$d was determined 
with about 1~GHz  accuracy, which corresponds to a relative
accuracy of 20~ppm and 5\% of the linewidth.

Evaluation of the most challenging systematic effect, the quantum
interference effect has been recently completed~\cite{Amaro2015}.
Due to the proximity  of two 2P states ($4\Gamma$ apart), 
the quantum interference effects might be considerable.
However, a quantitative evaluation of this effect, when accounting for
the used excitation and detection schemes, the detector geometry, the
laser direction and polarization etc.,
yields a line shift $\delta E \leq 0.001\Gamma$, thus far below the
statistical accuracy of our experiment.

Moreover, the theory in muonic deuterium has only recently
converged~\cite{Krauth2015} to a state which allows a precise
determination of the deuteron charge radius from the Lamb shift
measured in $\mu$d.
An impressive progress as been achieved in recent years both on the
``purely'' QED sector~\cite{Boriev7, Korzinin2013, Wundt2012}, as well
as in the computation of the TPE contribution~\cite{Hernandez2014,
  Pachucki2015, Carlson2014_deuterium}, yielding~\cite{Krauth2015}:
\begin{equation}
\Delta E_{\mu\mathrm{d}}(2S-2P_{1/2}) =  228.7766(10)
                     - 6.1102(3)\, r_d^2 
                     + 1.7091(200)\; \mathrm{meV}
\label{eq:deuterium}
\end{equation}
where the first term represent basically the ``pure'' bound-state QED
contributions, the second term with $r_d$ in fm the leading finite
size contribution (including mixed radiative-finite-size corrections)
and the third the TPE contribution.

Combining our measurements with the   prediction of
Eq.~(\ref{eq:deuterium}) we will obtain $r_d$ with a relative
accuracy of $u_r=4\times 10^{-4}$ limited by the TPE contribution.
A second route to a precise $r_d$ value is to combine the $r_p$
extracted from $\mu$p with the H-D isotopic measurements of the 1S-2S
transition~\cite{Parthey2011}.
A comparison of these two numbers will check the consistency of the muonic
results and will give new constraints to BSM theories, e.g. if
and how the "new force carrier" can couple to the neutron.


%
In 2013 and 2014 we have measured for the first time 2 transitions
frequencies in $\mu^4$He$^+$ and 3 in $\mu^3$He$^+$ 
with relative accuracy of about 40~ppm.
The uncertainty of the transition frequency measurement is entirely
given by statistical uncertainty, since systematics effects or
uncertainty related with the laser frequency calibration are $<2$~ppm.
To extract the nuclear charge radii from these measurements the
corresponding theoretical predictions have to be known. Preliminary
values read
%
%
\begin{eqnarray}
\Delta E_\mathrm{^4He}(2S-2P_{1/2}) & = & 1668.669(20)- 106.340\, r_\mathrm{^4He}^2 + 9.52(30)\; \mathrm{meV}\\
\Delta E_\mathrm{^3He}(2S-2P_{1/2}) & = & 1644.658(20)- 103.508\, r_\mathrm{^3He}^2 + 14.66(40)\; \mathrm{meV}\; .
\end{eqnarray}
 Final numbers can not be given at this stage because it requires
 sorting out of the several contributions calculated by various groups
 using different frameworks~\cite{Boriev7, Martynenko_He,
   Korzinin2013, Wundt2012}.
Similar to $\mu$d, in recent years there has been an impressive
progression of the TPE predictions using state-of-the-art nuclear
potentials, rendering the inelastic nuclear contribution with 5\%
accuracy~\cite{Ji13}.
Such an accuracy opens the way to alpha-particle and helion radii
determination with a relative accuracy better than $1\times 10^{-3}$
which will be compared with the very precise value available from
scattering~\cite{Sick2015-He}.
Still, an important contribution of the TFE in $\mu^4$He$^+$ and in
$\mu^3$He$^+$, related with the intrinsic nucleon polarizabilities, has
not yet been addressed by the community.
``Simple'' scaling as used in $\mu$d~\cite{Carlson2014_deuterium}
probably does not apply in this situation because of the smaller
separation between nuclear and nucleon energies.

Besides providing insights into the $r_p$ puzzle these nuclear
radii represent benchmarks  to check
few-nucleon ab-initio calculations, or vice versa to fix low-energy
coefficients (e.g. $c_D$ or $c_E$ of the three-nuclei-interaction)
describing the nuclear potential in effective field
theories~\cite{Piarulli2013}.
Moreover they can be used as anchor point for the $^6$He-$^4$He and
$^8$He-$^4$He isotopic shift measurements~\cite{Lu2013}.
The radii extracted from $\mu$He$^+$  measurements will be
used to disentangle the 4$\sigma$ discrepancy between two
$^3$He-$^4$He isotopic shifts measurements~\cite{Rooij, Pastor},
%
and their knowledge opens the way to enhanced bound-state QED
tests for one- and two-electrons systems in
``regular'' He$^+$~\cite{herrmann09} and He~\cite{kandula11}.

\section{Conclusions}
Various attempts have been made to find a solution of the proton
charge radius puzzle which has been exposed by the Lamb shift
measurement in muonic hydrogen.
A plethora of theoretical works has been devoted in refining and
rechecking the underlying theory necessary to extract the charge
radius from the muonic measurements, in proposing theories beyond the
standard model, in reanalyzing scattering data, and investigating
proton structure.
After all, the puzzle still persist. 
As a next step, new experimental inputs are required to provide
guidance.
Understanding of nuclear effects will be of primary importance for
the interpretation of the next muonic measurements.

\begin{acknowledgement}
This work is supported by the SNF\_200021L-138175, the
SNF\_200020\_159755, the DFG\_GR\_3172/9-1, the ERC StG. \#279765, by
the FCT, Portugal under Contract No. SFRH/BPD/76842/2011
SFRH/BPD/74775/2010, through the projects No. PEstOE/FIS/UI0303/2011,
PTDC/FIS/117606/2010, SFRH/BPD/92329/2013 and SFRH/BD/52332/2013.
We acknowledge fruitful discussions with
J.~Arrington, N.~Barnea, S.~Bacca, M.~Birse, E. Borie, C.~Carslon,
M.~Distler, M.~Gorchtein, C.~Ji, S.~Karshenboim, 
J.~McGovern, K.~Pachucki and S.~Schlimme.

\end{acknowledgement}

\end{document}